\documentstyle[prd,preprint,aps,preprint]{revtex}
\newcommand{\beq}    {\begin{equation}}
\newcommand{\eeq}    {\end{equation}}
\newcommand{\beqarr} {\begin{eqnarray}}
\newcommand{\eeqarr} {\end{eqnarray}}

\newcommand{\zc}     {\chi^{(0)}}
\newcommand{\fc}     {\chi^{(1)}}

\newcommand{\ovl}     {\overline}

\begin{document}
\draft
\preprint{
\begin{tabular}{r}
JHU-TIPAC 95026
\end{tabular}
}
\title{ A scale-dependent cosmology for the inhomogeneous Universe}
\author{C. W. Kim\footnote{E-mail cwkim@jhuvms.hcf.jhu.edu},
T.H. Lee\footnote{E-mail thlee@jhup.pha.jhu.edu}
and J. Song\footnote{E-mail jhsong@rowland.pha.jhu.edu}}
\address{* \dag \ddag Department of Physics and Astronomy\\
The Johns Hopkins University\\
Baltimore, MD 21218, U.S.A.\\
and}
\address{* \dag Department of Physcis, Soong Sil University\\
Seoul 156-743, Korea}

\maketitle

\begin{abstract}
\setlength{\baselineskip}{.5cm}
A scale-dependent cosmology is proposed
in which the Robertson-Walker metric and the Einstein
equation are modified in such a way that
$\Omega_0$, $H_0$ and the age of the Universe
all become scale-dependent.
Its implications on the observational cosmology are discussed.
\end{abstract}

\newpage
The standard Friedmann cosmology is based on the
Cosmological Principle (CP) which states that the Universe
is homogeneous and isotropic at any given time.
While isotropy has been well--established, for example,
from the observation of the Cosmic Microwave Background Radiation (CMBR)
by COBE \cite{COBE},
homogeneity has been challenged by various observations of
large scale structures such as filaments, sheets, superclusters,
voids and so on.
Moreover, one of the most remarkable consequences of recent
galaxy surveys\cite{pencil}\cite{ESP}
is that the scale of the largest structures
in each survey is comparable with the extent of the survey itself,
implying the absence of any tendency toward homogeneity
up to the present observational limit.
Instead, the observed matter density seems to be an increasing
function of scale from our underdense neighborhood\cite{Schramm}.
Another subject of heated controversy is
about the recent measurements of high values of $H_0$
\cite{Pierce}\cite{Freedman}
and their implied age of the Universe which becomes only half
the measured ages of $14 \sim 18$ Gyr for the oldest stars
and globular clusters.

The above situation has motivated us
to propose a cosmological model in which $\Omega_0$,
$H_0$ and the age of the Universe all become scale--dependent,
violating the homogeneity in CP.
Our model is based on the following two ansatzes.
First, we propose that the Universe is described by the metric
\begin{equation}
d\tau^2 = dt^2 -R^2(t,r)(dr^2 +r^2 d \Omega^2),
\end{equation}
where $R(t,r)$ is the generalized scale factor which
is a non--separable function of $r$ as well as $t$.
(If $R(t,r)$ is separable in the form of $a(t)f(r)$,
the Robertson--Walker metric, i.e., homogeneity is recovered.)
Since homogeneity in CP is violated in this metric,
only an observer at the center sees an isotropic Universe.
Based on the observations of isotropic CMBR and $\Omega(r)$,
we set our position at the center of the metric
for mathematical simplicity.
The next ansatz is to generalize the Einstein equation in two ways.
First, in order to accommodate the $r$--dependence of the Ricci tensors
calculated from the inhomogeneous metric (1),
the Einstein equation is generalized to
\begin{equation}
 R^{\mu \nu} - \frac{1}{2}g^{\mu \nu} R = - 8 \pi [GT^{\mu \nu}](t,r)~.
\end{equation}
Note that the energy--momentum conservation, $[GT^{\mu\nu}]_{;\nu}=0$,
is naturally obtained in view of the Bianchi identities.
Secondly, for an $inhomogeneous$ Universe,
a generalized stress--energy tensor with the contribution
of viscous fluid with heat conduct is introduced as
\begin{equation}
T^{\mu\nu}=\rho u^{\mu}u^{\nu}+(p-\zeta\Theta)P^{\mu\nu}
		+q^{\mu}u^{\nu}+u^{\mu} q^{\nu}~,
\end{equation}
where $\zeta \geq 0$ is the coefficient of $bulk$ viscosity,
 $\Theta=3\dot{R}/R$ and $P^{\mu\nu} \equiv
(u^{\mu}u^{\nu}-g^{\mu\nu})$ represent
the expansion and projection tensor of the fluid, respectively,
and $q^\mu$ is the heat--flux 4--vector with components
$q^{\mu}=(0,q^{(r)},0,0)$
in the local rest--frame of the $isotropic$ fluid.

Even though $p_r$ and $p_{\theta}(=p_{\phi})$
in Eqs.(1) and (2) appear to be different,
we assume $p_r=p_{\theta}$ to avoid shear forces,
yielding another constraint on $R$ as
\begin{equation}
\frac{R''(t,r)}{R^3(r,t)}-2\frac{R'(t,r)^2}{R^4(t,r)}
-\frac{R'(t,r)}{rR^3(r,t)} =0~~,
\end{equation}
which can be integrated, yielding
\begin{equation}
R(t,r)=\frac{a(t)}{1-\left[ {1+b(t) \over 4} \right]r^2}~~,
\end{equation}
where $a(t)$ and $b(t)$ are  arbitrary functions of $t$ alone.
In \cite{KS}, it is shown that setting $T^i_i=-(p-\zeta\Theta)$
and $T^{01}$ to be zero with Eq.(5), factorizes $R(t,r)$ as $a(t)f(r)$,
which is of no interest to us.
Physically, this non--zero $T^{01}$ has been responsible
for allowing matter  flow from its homogeneous distribution to the
present inhomogeneous one.
Now, we have two Einstein Field Equations (EFE)
\begin{equation}
\left[  \frac{\dot{R}(t,r)}{R(t,r)} \right]^2
=	\frac{8\pi }{3}[G\rho](t,r) + \frac{1}
{a^2(t)} + \frac{ b(t)}{a^2(t)}~~,
\end{equation}
\begin{equation}
\frac{\ddot{R}(t,r)}{R(t,r)}=-\frac{4 \pi}{3} [
	G\rho + 3G(p-\zeta\Theta) ]~~.
\end{equation}
In the local Universe($r\simeq 0$), $(\dot{R}/R)$ and
$(\ddot{R}/R)$ reduce, respectively, to $\dot{a}/a$ and $\ddot{a}/a$,
motivating us to interpret $a(t)$ as the scale--factor
of the local Universe with the modification $b(t)$.
Assuming the following form of the $\zeta\Theta$ term
\begin{equation}
8\pi G\zeta\Theta=
	\left[ \zc(t) \frac{\dot{a}_0}{a_0}+
	\fc(t) \frac{ \dot{R}}{R} \right]
		\frac{ \dot{R}}{R}~,
\end{equation}
the behaviors of $\rho$, $p$ and $T^{01}$ can be determined
by Eqs.(2) and (5) as
\begin{eqnarray}
\frac{8\pi}{3}G\rho &=&
	\left(\frac{\dot{a}}{a} \right)^2
	-\frac{1}{a^2}-\frac{ b}{a^2}
	+ 2\left(\frac{\dot{a}}{a}\right)D(t,r)
	+ D^2(t,r)	\\
8\pi Gp&=&
	\frac{1}{a^2}-
	2\frac{\ddot{a}}{a}
	+ \zc \frac{\dot{a}_0}{a_0} \frac{\dot{a}}{a}
	+[\fc -1] \left( \frac{\dot{a}}{a}\right)^2
	+\frac{b}{a^2}
	\\	\nonumber
	& &+ \left( \zc\frac{\dot{a}_0}{a_0}+
		(2\fc- 6) \frac{\dot{a}}{a}-
	2\frac{\ddot{b}}{\dot{b}} \right)D(t,r)
	+ [\fc-5] D^2(t,r)
			\\
8\pi GT^{01} &=& \frac{\dot{b}r}{a^2} ~~,
\end{eqnarray}
where $D(t,r)$ is given by
\begin{equation}
D(t,r) \equiv \left(
	\frac{\dot{b}r^2/4}
	{1- \left[ {1+b \over 4} \right]r^2} \right)~.
\end{equation}
{}From the observed $increase$ of the present mass density,
we assume $\dot{b}>0$ in the matter--dominated era.
The homogeneous Universe in the early radiation--dominated
era can be obtained in this model by requiring $\dot{b}=0$.
In the following, we restrict ourselves to the
matter-dominated era.
Now, the sign of $[1+b(t)]$ term becomes crucial
because of $D(t,r)$ terms in $\rho$ and $p$.
The spatial curvature, $R^{(3)}=-[1+b(t)]/a^2(t)$,
suggests that $[1+b]$ be positive in order
to explain the locally $open$ Universe,
implying an apparent singularity at $r=2/\sqrt{1+b}\equiv r_H$.
This singularity is, however, spurious one as $r=2GM$ singularity in
the Schwartzschild metric,
because $R^{(3)}$ is finite at $r=r_H$ as at $r=2GM$.
Therefore, the point $r=r_H$ is interpreted as an event horizon.

Let us discuss physics at this event horizon.
First, we have a specific relation between $p$ and $\rho$
at $r=r_H$. We have, from Eqs.(9) and (10),
\beq
\lim_{r \rightarrow r_H}
	\frac{Gp}{G\rho}=\frac{\fc(t)-5}{3} \equiv \gamma-1
{}~.
\eeq
The so--called gamma--law equation of state
 ($0\leq \gamma \leq 2$) can be satisfied by restricting the
value of $\fc$ in the range [2,8].
Since $(p-\zeta\Theta)$, not $p$, is involved in Eq.(7),
 dynamics at $r=r_H$ is described by
$[(p-\zeta\Theta)/\rho]_{r=r_H} \rightarrow -5/3$,
implying $\ddot{R}>0$ (see Eq.(7)).
This ($\ddot{R}>0$) is the very condition of the
generalized inflation \cite{Lucchin},
suggesting a picture of the Universe which is
inside an $expanding$ shell with infinite mass density.

In this model, cosmological quantities such as the
expansion rate, $\Omega$ and the age of the Universe are all
functions of $t$ and $r$.
The expansion rate and $\Omega$, in this model, are
\begin{equation}
H(t,r) \equiv \frac{\dot {R}(t,r)}{R(t,r)},
\end{equation}
\beq
\Omega(t,r) \equiv
\frac{\rho(t,r)}{\rho_c(t,r)}
=1-\frac{1+ b(t)}{a^2(t)}\frac{1}{ H^2(t,r)}~~,
\eeq
where $\rho_c$ is defined by $H^2\equiv 8\pi G\rho_c/3$.
We must caution the reader that
Eqs.(14) and (15) cannot be observed at $(t,r)$,
for every observation is based on the light propagation
given by
\beq
dr=-\frac{dt}{R(t,r)}~,
\eeq
whose solution is denoted by $r(t,t_{received})$,
explicitly showing the boundary condition
$r(t=t_{recieved})=0$.
In \cite{KLS}, Eq.(16) is numerically solved
with  the boundary condition $r(t_0,t_0)=0$
by specifying the functional forms of $a(t)$ and $b(t)$
under the assumption that $a(t)$ resembles the scale
factor of the Friedmann cosmology with $k=-1$
and $b(t)$ is a small modification.
Fitting them in the form of
$r(t,t_0)=\delta[t_0^{\kappa}-t^{\kappa}]$
gives an expression for the redshift $z$ as
$1+z=(t/t_0)^{\kappa -1}$.
With the definition of the luminosity distance $d_L$
as $d_L \equiv r R(t,r)(1+z)$,
the redshift--luminosity distance relation
is presented in \cite{KLS}.
The numerical calculations show that the linear relationship
between $z$ and $d_L$ for nearby objects ($z\ll 1$) remains intact
in this model also,
and the overall Hubble diagram for a locally $open$ Universe
with $\Omega(t_0, r\simeq 0)=0.1$ is similar to that of the $flat$
Friedmann cosmology.
What, then,  is the physical meaning of the observed increase
of $\Omega$ in this model?
Every direct information obtained by
light signal reaching us right now is
about the past, including $\rho(t)$, where $t<t_0$.
We deduce $\rho(t_0)$ from the observed $\rho(t)$
in the framework of the Friedmann cosmology, that is,
using $\rho(t)/\rho(t_0)=S^3(t_0)/S^3(t)=(1+z)^3$.
What is considered to be $\Omega_0$, therefore, is in this model,
\beq
\Omega^{obs}=\frac{\rho^{deduced}(t_0)}{\rho_{c,s}}
	=\frac{\rho(t,r(t))}{\rho_{c,s}(1+z)^3}~.
\eeq
With  the numerical solution described below Eq.(16),
$\Omega^{obs}(z)$ is plotted in \cite{KLS},
showing that $\Omega^{obs}(z)$  is indeed an increasing function
of $z$.

Finally, we discuss the age of the Universe,
which is in crisis in the framework of the Friedmann cosmology
due to the measurements of high values of $H_0$.
The age of the Universe in this model is also
$r$--dependent as $H(t,r)$ and $\Omega(t,r)$.
Since the information of $t_0(r)$ comes through
the light propagation with finite speed, however,
the measurement of the $present$ age of the Universe
at $r$ can only be accomplished in the future.
That is, only the $local$ age of the Universe is the
$true$ age of the Universe,
appropriate for comparison with the age of the stars in our galaxy.
Defining $x$ as $a \equiv a(t)/a(t_0)$,
Eq.(6) at $r \simeq 0$ yields
\beq
\dot{x} =\sqrt{
	\frac{8\pi G \rho(t,r\simeq 0)}{3}\frac{a^2(t)}{a^2(t_0)}
	+\frac{1+b(t)}{a^2(t_0)} }~.
\eeq
Assuming that the whole history of the Universe is
dominated by the matter--dominated era, when $T^i_i$ and $T^{01}$
are relatively small compared with $\rho$,
we have $\ovl{\rho}(t) \sim 1/a^3(t)$
from the energy--momentum conservation.
Then, Eq.(18) yields
\beq
\ovl{t}_0 \simeq \frac{1}{\ovl{H}_0}
	\int_0^1 \frac{\sqrt{x}}{
			\sqrt{ \ovl{\Omega}_0 +(1-\ovl{\Omega}_0)}}
	dx \simeq \frac{0.9}{\ovl{H}_0}~,
\eeq
where we have used Eq.(14) at $r\simeq 0$.
Note that the true age of the Universe is determined by the
$local$ values of $\Omega_0$ and $H_0$.
For example, with  $\ovl{\Omega}_0=0.1$ and $\ovl{H}_0=50$
Km/secMpc, the age of the Universe in this model is 18 Gyr,
which can easily accommodate the observed age $14 \sim 18$ Gyr
of the oldest stars and globular clusters in our galaxy.

In conclusion, the  scale--dependent cosmological
model proposed here can explain,
at least qualitatively,
the observed increase of the mass density,
the age problem and a conflict among current measurements of the
Hubble parameters.

\end{document}